\documentclass[submission,copyright,creativecommons]{eptcs}
\providecommand{\event}{FROM 2019} 
\usepackage{underscore}           

\usepackage{graphicx}
\usepackage{caption}
\usepackage{subcaption}
\usepackage{amsfonts,amstext}
\usepackage{amssymb,amsmath}
\usepackage{amsthm}

\newtheorem{property}{Property}

\newcommand{\class}[1]{\ensuremath{\mathrm{#1}}}

\newcommand{\Lang}{\ensuremath{\mathcal{L}}}
\newcommand{\Nat}{\ensuremath{\mathbb{N}}}

\newcommand{\mk}{\ensuremath{\mathbf{m}}}   
\newcommand{\In}[2]{\ensuremath{\mathrm{I}[#1,#2]}} 
\newcommand{\Out}[2]{\ensuremath{\mathrm{O}[#1,#2]}}
\newcommand{\Inh}[2]{\ensuremath{\mathrm{H}[#1,#2]}}

\newcommand{\Wplus}[2]{\ensuremath{\mathrm{W^+}[#1,#2]}}
\newcommand{\Wminus}[2]{\ensuremath{\mathrm{W^-}[#1,#2]}}

\newcommand{\AtP}[2]{\ensuremath{\mathrm{Ab}[#1,#2]}}
\newcommand{\SfP}[2]{\ensuremath{\mathrm{Rb}[#1,#2]}}

\newcommand{\SNnode}[1]{\ensuremath{\mathtt{#1}}}
\newcommand{\tuple}[1]{\ensuremath{\langle #1 \rangle}}

\title{An Operational Semantics of Graph Transformation Systems\\ Using Symmetric Nets}
\author{Lorenzo Capra
\institute{Dipartimento di Informatica\\
Universit\`a degli Studi di Milano\\
Milan, Italy}
\email{capra@di.unimi.it}
}
\def\titlerunning{An Operational Semantics of GTS}
\def\authorrunning{L. Capra}
\begin{document}
\maketitle

\begin{abstract}
Graph transformation systems (GTS) have been successfully proposed as a general, theoretically
sound model for concurrency. Petri nets (PN), on the other side, are a central and intuitive formalism for concurrent or distributed systems, well supported by a number of analysis techniques/tools. Some PN classes have been shown to be instances of GTS. In this paper, we change perspective presenting an operational semantics of GTS
in terms of Symmetric Nets, a well-known class of Coloured Petri nets featuring a structured syntax that outlines model symmetries.
Some practical exploitations of the proposed operational semantics are discussed. In particular, a recently developed structural calculus for SN is used to validate graph rewriting rules in a symbolic way.
\end{abstract}

\section{Introduction}
Graph transformation systems (GTS) are widely recognized as a general, well established formal model for concurrency. Petri nets (PN) \cite{ReisigPN}, on the other side, are a central model for concurrent or distributed systems.
Their success is due to several reasons, mostly, the fact that they can describe in a natural way the evolution of systems whose states have a distributed nature (this maps to the notion of PN \textit{marking}),
and the availability of a number of tools/techniques supporting the editing/analysis of PN models.

Petri nets are a reference model for any formalism meant to describe concurrent or distributed systems, including GTS. It is well known that GTS are a generalization of some PN classes, as shown by Kreowsky in its pioneering work \cite{Kre80} using the double-pushout approach. Basically, the idea is to represent a marked PN as a graph with three different types of nodes (for places, transitions, and tokens) and describe the firing of a PN transition thorough a rule (derivation). Since then, several encodings of PN classes in terms of GTS have been presented, among which Place/Transitions nets, Condition/Event nets, Elementary Net Systems, Consume-Produce-Read nets.
Some net variants with extra features such as read/reset/inhibitor arcs have been also encoded. It is impossible to exhaustively list all these proposals,
let us refer to \cite{Cor96} (and included references) for the earliest and
\cite{BaCoGaMon},\cite{Ehrig03} for more recent ones.

In this paper we consider the relationship between GTS and PN from a new perspective: we provide a formalization of Graph Transformation Systems (GTS) based on Symmetric Nets (SN)\footnote{formerly known as Well-formed Nets, or WN} \cite{ARCDFH93}, a type of Coloured Petri nets \cite{HLPN},\cite{JensenBook1} featuring a particular syntax that outlines model symmetries and is exploited both in state-space based and structural analysis. The idea is simple: each rule (derivation) of a GTS corresponds to a SN transition which is properly connected to a couple of SN places whose marking encodes a graph. 
In the paper we refer to simple directed graphs, even if the approach might be generalized to any category of (hyper)graphs.

The advantages of this approach are numerous, and the aim of the paper is to illustrate some of them though some examples: we can exploit well established tools supporting the editing/analysis of SN, like the GreatSPN package
\cite{Baarir:09}; an operational interleaving semantics for GTS is provided in a natural way building the state-transition system of a SN; a compact state-transition system -called symbolic reachability graph \cite{CHIOLA97}, in which states (markings) representing isomorphic graphs are folded, can be directly derived once an initial symbolic graph encoding is set; some recent advances in SN (symbolic) structural analysis \cite{ATPN2005}, \cite{QEST2015},
implemented in the SNExpression tool (\url{www.di.unito.it/~depierro/SNex}) may be exploited to check some conditions ensuring rule well-definiteness, validate rules, and verify their potential concurrency;
in particular, a fully automated calculus of symbolic structural relations in SN models may be profitably used. All these concepts are instantiated on a few, though significant, examples of graph rewriting rules, and a simple GTS.
All the examples used in the paper are available in GreatSPN format at
\url{https://github.com/lgcapra/GTS-SN}.

The GTS formalization based on SN may be considered as an alternative to classical approaches, in particular the algebraic ones based on single and double pushout. The strengths of this new proposal are a more intuitive definition of derivations, and the availability of well established tools for the editing/validation/analysis of models. The relationship between SN rules and single/double pushout derivations, however,
is not treated in this paper, and deserves further investigations.   

The balance of the paper is as follows:
Section \ref{sec:background} introduces SN and related background notions; Section \ref{sec:SNandGTS} presents the encoding of a GTS as a SN, and its operational semantics; symbolic structural conditions for rule well-definiteness are also set up; Section \ref{sec:struct-an} shows an application of SN structural calculus for verifying rule concurrency in a GTS;
finally, Section \ref{sec:concl} contains the conclusion and describes ongoing work

\section{Symmetric Nets}
\label{sec:background}

In this section we present the SN formalism and a few preliminary concepts and notations used in the sequel.
We let the reader refer to~\cite{ReisigPN} and~\cite{ARCDFH93} for a complete treatment of Petri nets and SNs, respectively. 

\subsection{Multisets}
A multiset (or \textit{bag}) over a domain $D$ is a map $b: D \rightarrow \Nat$, where $b(d)$ is the \emph{multiplicity} of $d$ in $b$. The \textit{support} $\overline{b}$ is the set $\{d \in D | b(d) > 0 \}$: we write $d \in b$ to mean $d \in \overline{b}$. 
A multiset $b$ may be denoted by a weighted formal sum of $\overline{b}$ elements where coefficients represent multiplicities. The \textit{null} multiset (over a given domain), i.e., the multiset with an empty support, is denoted (with some overloading) $\emptyset$.
The set of all bags over $D$ is denoted $Bag[D]$.
Let $b_1,b_2 \in Bag[D]$. The sum $(b_1+b_2) \in Bag[D]$ and the difference $(b_1-b_2) \in Bag[D]$ are defined as:
$(b_1+b_2)(d) = b_1(d) + b_2(d)$; $(b_1-b_2)(d) = max(0,b_1(d)-b_2(d))$.  Also relational operators are defined component-wise, e.g., $b_1 < b_2$ if and only if $\forall d$, $b_1(d) < b_2(d)$.\\
The {\it scalar} product
$k \cdot b_1$, $k \in \mathbb{N}$, is $b_1^\prime \in Bag[D]$, s.t. $b_1^\prime(d) = k \cdot b_1(d)$. Let  $b_1 \in Bag[A]$, $b_2 \in Bag[B]$, and so forth:
the {\it Cartesian} product $b_1 \times b_2  \times \ldots \in Bag[A \times B \times \ldots]$ is
defined as $(b_1 \times b_2 \times \ldots) (\tuple{a,b,\ldots}) = b_1(a) \cdot b_2(b) \cdot \ldots$

\paragraph{Multiset functions} All the operators on multisets straightforwardly extend to functions mapping to multisets.
Let $f_1, f_2: D \rightarrow Bag[D']$; if $op$ is a binary operator on bags, then $f_1 \ op \ f_2$ is defined as $f_1 \ op \ f_2 \ (a) = f_1(a) \ op \ f_2(a)$. Analogously if $op$ is a unary operator: e.g., $\overline{f_1}$  is a function $D \rightarrow 2^{D'}$ such that  $\overline{f_1}(a) = \overline{f_1(a)}$. As for relational operators, $f_1 < f_2$ if and only if $\forall a, f_1(a) < f_2(a)$. With some overloading, the symbol $\emptyset$ will denote a constant null multiset function.\\
Let $f_1: \ D \rightarrow Bag[A]$, $f_2: \ D \rightarrow Bag[B]$, and so forth: the product $f_1 \times f_2 \times \ldots \ : D \rightarrow Bag[A \times B \times \ldots]$ is defined: $f_1 \times f_2 \times \ldots (d) \ = f_1(d) \times f_2(d) \times\ldots$. In the following a function-tuple $\tuple{f_1,f_2,\ldots}$ will denote the function Cartesian product $f_1 \times f_2 \times \ldots$\\
Let $f: D \rightarrow Bag[D']$: the \textit{transpose}  $f^t$ $:D' \rightarrow Bag[D]$ is defined as: $f^t(x)(y) = f(y)(x), \forall x \in D', y \in D$;
the \textit{linear extension}
$f^*: \ Bag[D] \rightarrow Bag[D']$ is defined as $f^*(b) = \sum_{x \in b} b(x) \cdot f(x)$.
The composition operator is extended accordingly:
let $h: \ A \rightarrow Bag[B]$, $g: \ B \rightarrow Bag[C]$,
then $g \circ h:$ $A \rightarrow Bag[C]$ is defined as $g \circ h (a) = g^* (h (a))$. For simplicity, we will use the same symbol for a function and its linear extension.

\subsection{Symmetric Nets}
\label{sec:SNform}

Symmetric Nets (SN) \footnote{Introduced with the name of Well-formed Nets, later renamed SNs.}~\cite{ARCDFH93} are a high-level Petri Net formalism featuring a particular syntax for places, transitions, and arc annotations: such syntax has been devised to make the symmetries present in model's structure and behaviour explicit. This formalism is thus convenient from the point of view of model representation as well as from that of its analysis. Efficient methods have been proposed to perform SN state-space based analysis~\cite{CHIOLA97}, or structural analysis~\cite{ATPN2005},\cite{QEST2015}. Many of these algorithms have been implemented in GreatSPN~\cite{Baarir:09}, whereas  the most recent developments on structural analysis have been implemented in SNexpression  (\url{www.di.unito.it/~depierro/SNex}).

SN are a particular flavour of \emph{Colored} Petri nets (PN), originally introduced in \cite{JensenBook1}.
Like in any Petri net, the SN underlying structure is a kind of (finite) directed bipartite graph, where the set of nodes is $P \cup T$, $P$ and $T$ being non-empty, disjoint sets, whose elements are called \textit{places} and \textit{transitions}, drawn as circles and bars, respectively. The former represent system state variables, whereas the latter events causing (local) state changes:
what characterizes Petri nets in fact is a distributed notion of state, called \textit{marking}. As in any high-level PN model, both places and transitions are associated with (colour) domains. Edges are annotated by (colour) functions mapping the domain of the incident transition to the domain of the incident place.

This section introduces the SN formalism exemplifying some key concepts by means of the models used in the rest.

\subsubsection{Colour Domains}
SN \emph{places} are associated with a \textit{color domain} ($cd$) defining the type
of tokens a place may hold. A color domains is a Cartesian product of finite, non-empty, pair-wise disjoint \emph{basic color classes},
denoted by capital letters (e.g., \class{C}).
Basic color classes may be partitioned into \emph{static subclasses} (denoted by capital letters with a subscript, e.g., $\class{C}_1$), or, in alternative, \emph{circularly} ordered.

The SN models defined later build on a single basic color class: \class{N}=$\{nd_i\}$. The place color domains are \class{N} and \class{E} = $\class{N} \times \class{N}$ (or $\class{N}^2)$.

\emph{Transitions} have a color domain as well, since they specify parametric events. The color domain of a transition is implicitly determined by transition's parameters (\emph{variables}) that annotate incident edges and transition's guard, denoted in this paper by lower-case letters with a subscript, e.g. $c_i$. By convention, the letter used for a variable implicitly defines its type, i.e., the color class denoted by the corresponding capital letter. Subscripts are used to distinguish variables of a given type associated with a transition.
As an example, the colour domain of transition \SNnode{R1} (Figure \ref{fig:R1}) is
$\class{N} \times \class{N} \times \class{N}$.

If no variable symbols surround a given transition, its domain is implicitly defined by  a singleton \textit{neutral color}.

\subsubsection{Transition guards}
Transitions may have \emph{guards}, consisting of \emph{boolean predicates} defined on transition domains:
\begin{itemize}
\item $[c_1= (\neq)c_2]$ is true when the same/a different color is assigned to $c_1$ and $c_2$;
\item $[c_1 \in \class{C}_j ]$ is true when the color assigned to $c_1$ belongs to subclass $\class{C}_j$; 
\item $[d(c_1)=d(c_2)]$ is true when the colors assigned to $c_1$ and $c_2$  belong to the same subclass.
\end{itemize}

A \emph{transition instance} is a pair $(t,b)$, where $b$ (\emph{binding}) is an assignment of colors to the transition's variables.
For instance, a possible binding for \SNnode{R1} is $n_1 = nd_2$, $n_2 = nd_1$,  $n_3 = nd_3$. A transition instance is \emph{valid} if it satisfies the transition's guard. From now on with transition color domain we will mean the set of valid transition instances.

A transition guard is omitted if and only if it is the constant $true$.

\subsubsection{Marking} A \textit{marking} \mk\ provides a distributed notion of system state. Formally, a marking maps every place to a multiset on its domain:
$\mk(p) \in Bag[cd(p)]$ is the marking of place $p$.
The elements of one such a multiset are called \textit{tokens}.

\subsubsection{Arc Functions}
An arc form a place $p$ to a transition $t$ is called \textit{input} arc, whereas one in the opposite direction is called \textit{output} arc. A place and a transition may be also connected by an \textit{inhibitor arc}, drawn with an ending small circle instead of an arrow. Arcs are annotated by corresponding \emph{arc functions},  
denoted by \In{p}{t}, \Out{p}{t} and \Inh{p}{t}, respectively.
An arc function is a map
$F:\,cd(t) \rightarrow Bag[cd(p)]$, formally expressed as a linear combination:
\begin{equation}
\label{eq:SNarcfunction}
F = \sum_i \lambda_i . T_i,~ \lambda_i \in \Nat,
\end{equation}
where $T_i$ is a tuple (i.e., a Cartesian product) of \emph{class functions} $\langle f_1, \ldots , f_k \rangle$.

A class-$\mathrm{C}$ function $f_i$ is a map $cd(t) \rightarrow Bag[\class{C}]$, expressed in turn as a linear combination of functions in an elementary set:
\begin{equation}
\label{classfun}
f_i = \sum_{h} \alpha_h. e_h,~ \alpha_h\in {\mathbb Z},
\end{equation}
where (referring to class \class{C}) $e_h \in \{c_j,{\scriptstyle ++}c_j,\mathrm{C}_q, All\}$:
\begin{itemize}
\item $c_j$ (previously called variable) is actually a \emph{projection}, i.e, given a tuple of colours in $cd(t)$ maps to the $j^{th}$ occurrence of color $\mathrm{C}$;
if class $\mathrm{C}$ is ordered, then ${\scriptstyle ++}c_j$ denotes the successor $\mathrm{mod}_{|\mathrm{C}|}$ of the element that $c_j$ maps to;
\item $\mathrm{C}_q$ and $All$ are \emph{diffusion} (or constant) functions mapping any color in $cd(t)$ to $\sum_{x \in \class{C}_q} 1 \cdot x$ and $\sum_{x \in \class{C}} 1 \cdot x$, respectively.
\end{itemize}
Scalars $\alpha_h$ in (\ref{classfun}) must be such that no negative coefficients result from the evaluation of $f_i$ for any legal binding of $t$.
Both function-tuples and class-functions may be suffixed by a guard defined on $cd(t)$, acting as a filter: $f[g](a) = f(a)$ if $g(a)$, otherwise $f[g](a) = \emptyset$.
If $t$ has an associated guard $g(t)$ then we assume $g(t)$ implicitly spans over all surrounding arc functions.

AS an example of arc function, consider the function on the
inhibitor arc connecting transition \SNnode{R2} to place \SNnode{Edge} (Figure \ref{fig:R2}). The transition's domain is $cd(\SNnode{R2}) = N$, because only variable $n_1$ occurs in incident edges. The evaluation of this function on a given $nd_i \in N$ results in the (multi)set composed of all pairs with the first element equal to $nd_i$ and all pairs
with the 2nd element equal to $nd_i$ and the first one other than $nd_i$.

The only basic class used in the SN models of the paper is neither partitioned nor ordered. Arc functions, moreover, map to multisets with multiplicities $\leq 1$. i.e., sets.

\subsubsection{SN Execution}
The interleaving semantics of a SN is fully defined by the \textit{firing rule}.
Assuming that missing arcs
(of any type) between SN nodes are arcs annotated by the null function $\emptyset$,
an instance $(t,b)$ is \textit{enabled} in marking \mk\ iff:
\begin{itemize}
\item $\forall p \in P$: $\In{p}{t}(b) \leq \mk(p)$
\item $\forall p \in P$, $x \in \Inh{p}{t}(b)$: $\Inh{p}{t}(b)(x) > \mk(p)(x)$ 
\end{itemize}

An instance $(t,b)$ enabled in \mk\ may \emph{fire} by withdrawing from each input place $p$
the bag $\In{p}{t}(b)$ and adding to each output place $p$ the bag $\Out{p}{t}(b)$. We get a new marking $\mk'$, formally defined as:

$$\forall p: \, \mk'(p) = \mk(p) - \In{p}{t}(b) + \Out{p}{t}(b) $$

\noindent We say that $\mk'$ is \textit{reachable} from $\mk$ through $(t,b)$, and this is denoted $\mk [ t,b > \mk'$.

Once an \textit{initial marking} $\mk_0$ of a SN is set,
it is possible to build the state-transition system (often called   \textit{reachability graph}, or RG) describing a SN model's behaviour.
The RG is a (edge-labelled) directed multi-graph inductively defined as follows: $\mk_0 \in RG$; if $\mk \in RG$, and $\mk [ t,b > \mk'$, also $\mk' \in RG$ and there is an edge $\tuple{\mk ,\mk'}$ with label $(t,b)$.

If a \textit{symbolic} initial marking is set, a quotient graph called \textit{symbolic reachability graph} is directly built, that retains all the information of the ordinary reachability graph. We will get to that later.

\section{Encoding GTS in SN}
\label{sec:SNandGTS}

In this section we show how to encode a Graph Transformation Systems through Symmetric nets.
Graph rewriting \textit{rules} are formalized in terms of SN transitions connected to a couple of shared places. They will be illustrated by a few examples.
For the sake of simplicity we refer to simple directed graphs, even if this approach may be extended to any category of (hyper)graphs.

A \textit{directed graph} (form now on simply graph) is composed of a (finite) set $N$ of nodes and a set $E \subseteq N \times N$ of edges. A (total) \textit{morphism} between graphs
$G_1 = (N_1,E_1)$ and $G_2 = (N_2,E_2)$ is a pair of functions $f_E : E_1 \rightarrow E_2$, $f_N : N_1 \rightarrow N_2$ such that $\forall \tuple{n_1,n_2} \in E_1$,
$f_E(\tuple{n_1,n_2}) = \tuple{F_N(n_1),F_N(n_2)}$.

\subsection{Graph encoding}
\label{sec:graphenc}
The graph encoding through SN builds on a couple of places, \SNnode{Node} and \SNnode{Edge}, whose associated colour domain are the basic colour class $\class{N} = \{nd_i \}$, and the product $\class{E} = \class{N} \times \class{N}$, respectively. We assume that class \class{N} holds enough elements to cover all possible evolutions of a graph.

A graph $G_1 = (N_1,E_1)$ is straightforwardly encoded by a SN marking, denoted $\mk_{G_1}$: letting $l$ be an injective labelling $N_1 \rightarrow \class{N}$, $\mk_{G_1}(\SNnode{Node}) = \sum_{n \in N_1} 1\cdot l(n)$, $\mk_{G_1}(\SNnode{Edge}) = \sum_{\tuple{n_1,n_2} \in E_1} 1\cdot \langle l(n_1), l(n_2) \rangle$.

The other way round, a SN marking \mk\ is a \textit{graph-encoding} if and only if both $\mk(\SNnode{Node})$ and
$\mk(\SNnode{Edge})$ are multisets whose elements have multiplicities $\leq 1$ (i.e., sets) and any colour $nd_i$ occurring in $\mk(\SNnode{Edge})$ also occurs in $\mk(\SNnode{Node})$ (there are no dangling edges).

\subsection{Graph rewriting rules}
\label{sec:graphrw}
A graph rewriting rule (or derivation) is formalized by a SN transition
$\SNnode{R}_i$ properly connected to places \SNnode{Node} and \SNnode{Edge}. The colour domain of $\SNnode{R}_i$ depends on how many variables (projections) $n_i$ occur on the incident arcs and transition's guard: in general, $cd(\SNnode{R}_i) = \class{N}^k$,
$k > 0$.

The idea is simple: the input arc functions $\In{\SNnode{Node}}{\SNnode{R}_i}$, $\In{\SNnode{Edge}}{\SNnode{R}_i}$ (assumed non both null), and the inhibitor arc function  $\Inh{\SNnode{Edge}}{\SNnode{R}_i}$, when evaluated on an enabled instance of $\SNnode{R}_i$ in a graph-encoding marking \mk, match
a subgraph of the encoded graph which is rewritten according to the SN firing rule: the matched subgraph is atomically removed from the encoded graph and replaced with the subgraph yielded by evaluating the output arc functions on the same instance. Inhibitor arc functions, even if not directly involved in the firing, play a crucial role both in the matching step and in setting structural conditions for rule correctness, as explained below.

\begin{figure}
\centering
\begin{subfigure}{.49\textwidth}
  \centering
  \includegraphics[width=\linewidth]{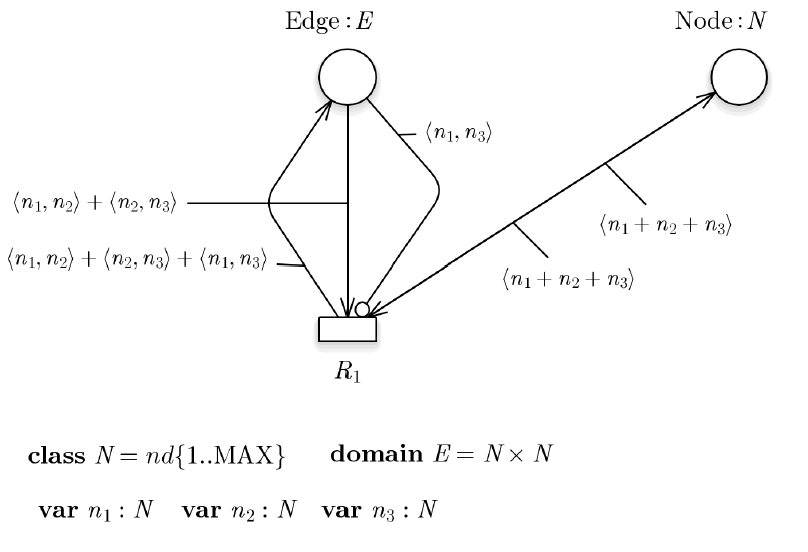}
  \caption{Rule 1}
  \label{fig:R1}
\end{subfigure}%
\begin{subfigure}{.49\textwidth}
  \centering
  \includegraphics[width=\linewidth]{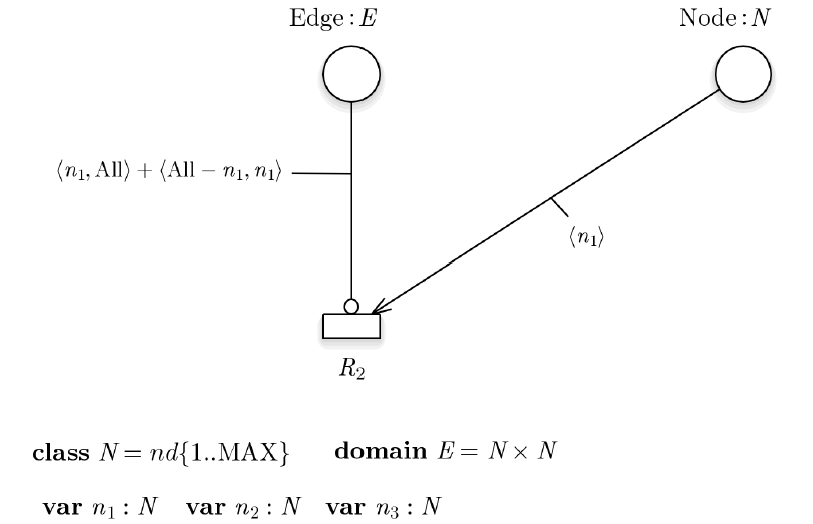}
  \caption{Rule 2}
  \label{fig:R2}
\end{subfigure}

\begin{subfigure}{.49\textwidth}
  \centering
  \includegraphics[width=\linewidth]{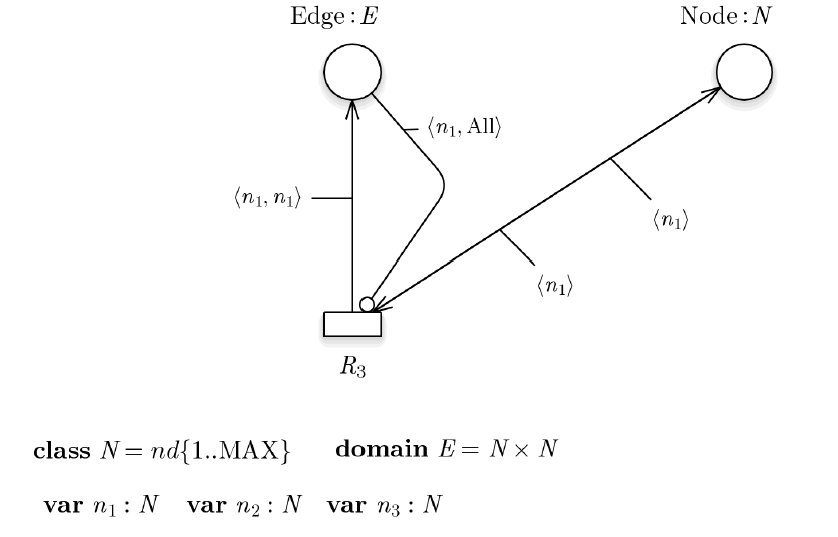}
  \caption{Rule 3}
  \label{fig:R3}
\end{subfigure}%
\begin{subfigure}{.49\textwidth}
  \centering
  \includegraphics[width=\linewidth]{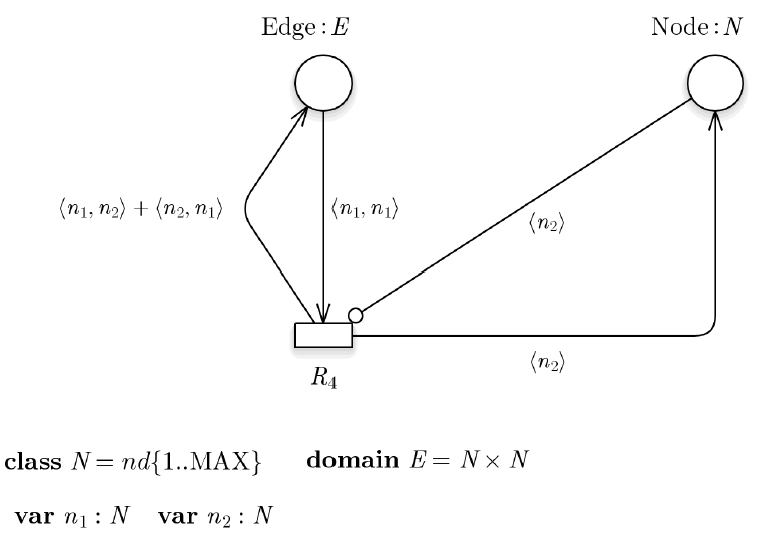}
  \caption{Rule 4}
  \label{fig:R4}
\end{subfigure}
\begin{subfigure}{.52\textwidth}
  \centering
  \includegraphics[width=\linewidth]{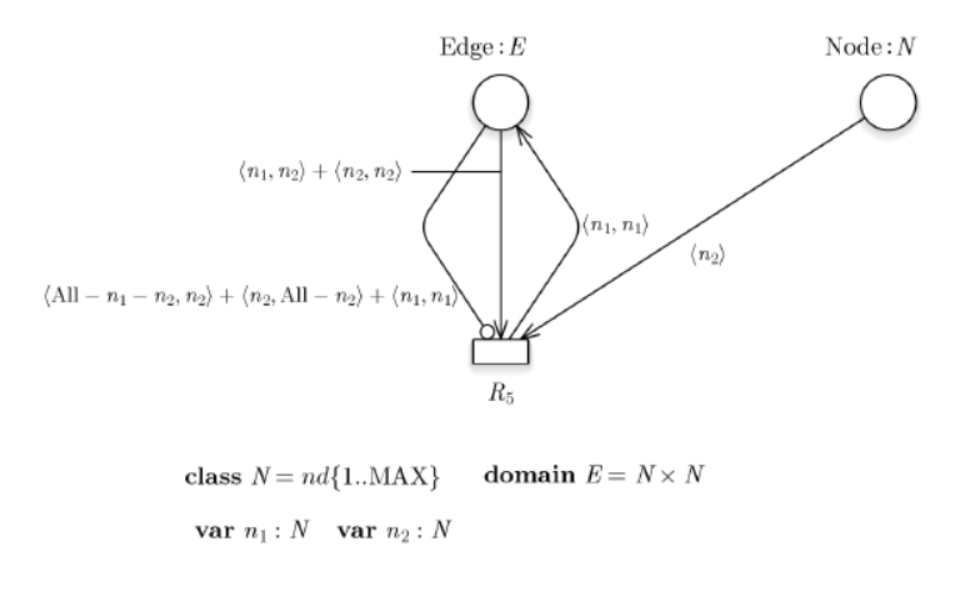}
  \caption{Rule 5}
  \label{fig:R5}
\end{subfigure}%
\begin{subfigure}{.47\textwidth}
  \centering
  \includegraphics[width=\linewidth]{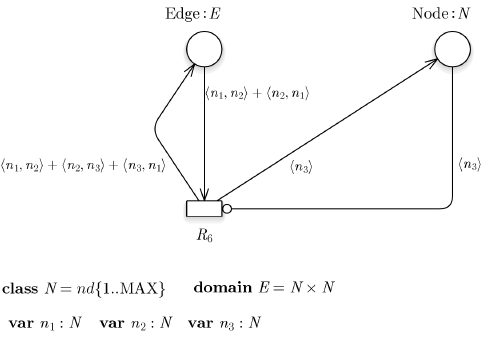}
  \caption{Rule 6}
  \label{fig:R6}
\end{subfigure}
\caption{Examples of graph rewriting rules}
\label{fig:rules}
\end{figure}

Some representative examples of rules are shown in Figure \ref{fig:rules}. Rule \ref{fig:R1} allows the transitive closure of a graph be incrementally computed. Rule \ref{fig:R2} represents
the removal of isolated nodes of a graph. Rule \ref{fig:R3} may be used to derive a Kripke structure from a graph: in fact, a self-loop is created for nodes without successors. Rule \ref{fig:R4} transforms a self-loop involving node $nd_i$ into a pair
of edges from/to $nd_j$, where $nd_j$ is a new node.
Rule \ref{fig:R5} is matched by a node $nd_i$ having as only successor $nd_j$, which has no other link but a self-loop: in that case $nd_j$ is removed, and a self-loop involving $nd_i$ is created. Finally, Rule \ref{fig:R6} translates a loop between $nd_i$ and $nd_j$
into a loop involving these two nodes and a newly inserted one.

\subsection{Well defined Rules}
\label{sec:rulecorr}

We have to establish some conditions ensuring that a rewriting rule is \textit{well-defined}, that is, \textit{any} instance of the rule (transition)
rewrites a (simple) directed graph into another one.
By exploiting the calculus for SN introduced in \cite{QEST2015}, \cite{ATPN2005}, it is possible to characterize these rules as structural conditions on the arc functions annotating the corresponding transition, that may be checked in a fully  symbolic and automated way, e.g., by using the SNexpression  (\url{www.di.unito.it/~depierro/SNex}) toolset.

The calculus for SN has been developed to check basic structural properties (conflict, causal connection, mutual exclusion) on SN without any net unfolding.
It builds on
the ability to solve in a symbolic way expressions whose terms are the elements of a language \Lang\, and involving a specific set
of functional operators (in this context, the difference, the composition, and the support).
The terms of \Lang\ are a small extension of the SN arc
functions, but the language restriction used here exactly matches SN arc functions. The calculus has been implemented as a rewriting system that, given any structural expression, reduces it to a normal form in \Lang. In particular, if $e \equiv \emptyset$ then $e \rightarrow \emptyset$.

In the following, the expressions \Wplus{p}{t} and \Wminus{p}{t} stand for \Out{p}{t}$-$\In{p}{t} and \In{p}{t}$-$\Out{p}{t}, respectively: they map any transition instance $(t,b)$ to the (multi)set of coloured tokens that (upon its firing) are added/withdrawn to/from place $p$.

Two type of terms are used: functions mapping to multisets, and their supports, mapping to sets. According to the type of operands, '$-$','$+$' will denote the multiset difference/sum or the set difference/sum. The same for the Cartesian product.
These equivalences are exploited (with an obvious overloading of symbol '$\emptyset$'):
\begin{center}
$F \leq G$ $\Leftrightarrow$ $ F - G \equiv \emptyset$; $\overline{F} \subseteq \overline{G}$ $\Leftrightarrow$ $ \overline{F} - \overline{G} \equiv \emptyset$.
\end{center}

Let \SNnode{R} be the transition encoding a rule. The conditions below ensure that \SNnode{R} is well defined:
\begin{align*}
1) \quad &  \Inh{\SNnode{Edge}}{\SNnode{R}}  \leq \tuple{All,All} \ \ \wedge \ \  \Inh{\SNnode{Node}}{\SNnode{R}}  \leq \tuple{All} \\
2) \quad & \Wplus{\SNnode{Edge}}{\SNnode{R}} \leq \tuple{All,All} \\
3) \quad &   \Wplus{\SNnode{Node}}{\SNnode{R}} \leq \Inh{\SNnode{Node}}{\SNnode{R}} & \\
4) \quad  & \mathrm{let} \ \ NA = (\overline{\tuple{n_1 + n_2} \circ \Out{\SNnode{Edge}}{\SNnode{R}}} - \overline{\tuple{n_1 + n_2} \circ \In{\SNnode{Edge}}{\SNnode{R}}}) - \overline{\In{\SNnode{Node}}{\SNnode{R}}}: \ \ NA \subseteq \overline{\Out{\SNnode{Node}}{\SNnode{R}}}  & \\
5) \quad & \overline{\Wplus{\SNnode{Edge}}{\SNnode{R}}} -
(\tuple{NA,\overline{All}} + \tuple{\overline{All},NA}) \subseteq \overline{\Inh{\SNnode{Edge}}{\SNnode{R}}} \\
6) \quad & \overline{((\tuple{All-n_1,n_1}+\tuple{n_1,All}) \circ   \Wminus{\SNnode{Node}}{\SNnode{R}})} - \overline{\Wminus{\SNnode{Edge}}{\SNnode{R}}} \subseteq \overline{\Inh{\SNnode{Edge}}{\SNnode{R}}}  &
\end{align*}

Conditions 1,2) are related to simplicity (these conditions alone, however, doesn't ensure it);
1) means that inhibitor arc functions map to multisets with multiplicities $\leq 1$, i.e., we can only check for the \textit{absence} of nodes/edges in a graph-encoding; 
2) means that new edges are inserted with multiplicity 1;
3) avoids node duplication.
Conditions 4-6) avoid (among others)
the creation of dangling edges, and are a bit more complex, involving the composition operator: 
4) means that the nodes incident to newly added edges, but that do not exist yet (this set of nodes is denoted $NA$), must be contextually inserted: it builds on the assumption that, in the current graph encoding, there are no dangling edges; 
5) is related, again, to simplicity: whenever a new edge is added, we must check its absence unless one of its incident nodes belongs to the precomputed set $NA$;
finally, 6) deals with node removal: the inhibitor arc function must ensure that,
for every withdrawn node, there are no edges incident to it, but for those edges that are contextually removed by the rule.

A few remarks have to be done. In condition 4), the domain of projections $n_1,n_2$ is $\class{N} \times \class{N}$, whereas in 6) the domain of $n_1$ is  $\class{N}$. The use of support operator in 4-6) is due to the fact that a composition may result in ordinary multisets, with multiplicities greater than one. The \textit{parametric} set $NA$ is computed by separately considering the output and the input arc functions to/from place \SNnode{Edge}, instead of considering \Wplus{\SNnode{Edge}}{\SNnode{R}}: in fact,
$\overline{\tuple{n_1 + n_2} \circ \Out{\SNnode{Edge}}{\SNnode{R}}} - \overline{\tuple{n_1 + n_2} \circ \In{\SNnode{Edge}}{\SNnode{R}}} \subseteq \overline{\tuple{n_1 + n_2} \circ \Wplus{\SNnode{Edge}}{\SNnode{R}}}$, therefore the condition we set is more general.

\begin{property}
If a rule/transition \SNnode{R} meets conditions 1-6),
then the firing of any instance (\SNnode{R},b) in a graph-encoding marking generates a graph-encoding marking.
\end{property}

\noindent The proof is just a direct consequence of the explanation above.
We can easily check that all rules shown in Figure \ref{fig:rules} 
are well defined.

\subsection{Bringing rules together}
\label{sec:tran-syst}

A Graph Transformation System (or GTS) may be very simply defined by bringing together a set of well-defined rules (transitions) sharing places \SNnode{Node} and \SNnode{Edge}, and setting an initial graph-encoding marking. The induced state-transition system corresponds to the SN reachability graph.

\begin{figure}
\centering
\includegraphics[width=\linewidth]{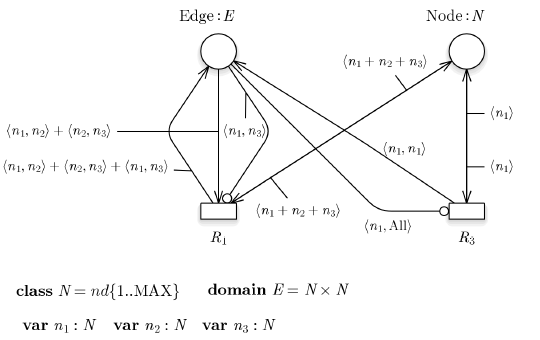}
\caption{a simple GTS composed of Rules 1,3}
\label{fig:R1-3}
\end{figure}

As an example, 
consider the SN in Figure \ref{fig:R1-3}. It comes from the combination of Rules 1,3) described above. Given a graph $G_0$ encoded by the initial marling $\mk_{G_0}$,
the derived RG describes the sequence of transformations that $G_0$ undergoes by applying either Rule 1 or Rule 3. The resulting RG has an \textit{absorbing} state, i.e. a dead home-state, which corresponds to the transitive closure of $\mk_{G_0}$ where nodes without proper predecessors are sources/targets of self-loops.

Let $\mk_{G_0}(\SNnode{Edge})= \tuple{nd_1,nd_2}+\tuple{nd_1,nd_3}+\tuple{nd_4,nd_1}$, and $\mk_{G_0}(\SNnode{Node} )= \tuple{nd_1+nd_2+nd_3+nd_4}$: the corresponding RG (built with the GreatSPN package) holds 16 nodes, one of which absorbing; this final node encodes the graph
$$\tuple{nd_1,nd_2}+\tuple{nd_1,nd_3}+\tuple{nd_4,nd_1}+\tuple{nd_2,nd_2}+\tuple{nd_3,nd_3}+\tuple{nd_4,nd_2}+\tuple{nd_4,nd_3}$$.
\paragraph{A Symbolic State-transition System}
During the construction of the SN reachability graph some markings encoding \textit{isomorphic} graphs may be reached. Consider the example above: from the initial marking, we can reach the two markings below \footnote{we refer to place \SNnode{Edge}, because the marking of \SNnode{Node} doesn't change} by firing $\SNnode{R}_1$ with the bindings
 $n_1 = nd_4$, $n_2 = nd_1$, $n_3 = nd_2$ and
$n_1 = nd_4$, $n_2 = nd_1$, $n_3 = nd_3$, respectively:
$$i) \ \tuple{nd_1,nd_2}+\tuple{nd_1,nd_3}+\tuple{nd_4,nd_1}+\tuple{nd_4,nd_2} \ \
ii) \ \tuple{nd_1,nd_2}+\tuple{nd_1,nd_3}+\tuple{nd_4,nd_1}+\tuple{nd_4,nd_3}$$

Observe that $i)$ and $ii)$ are isomorphic since can be obtained from one another by swapping $nd_2$ with $nd_3$. Recognizing isomorphic graph-encodings is for free in SN, if the initial marking is \textit{symbolic}.
A \textit{symbolic marking} $\widehat{\mk}$ \cite{CHIOLA97} is an equivalence class of ordinary markings: $\{ \mk_1, \mk_2 \} \subseteq \widehat{\mk}$ if and only if they correspond, up to a \textit{permutation} on colour classes (preserving the possible partitions in subclasses)

A symbolic marking (or SM) is syntactically expressed using \textit{dynamic subclasses} instead of ordinary colours. Dynamic subclasses define \textit{parametric partitions} of basic colour classes: each dynamic subclass is associated with a colour class (or a static subclass, if the class is split) and has a cardinality.
As an example, the initial symbolic marking encoding (among others) graph $G_0$ above is:
\begin{center}
$\widehat{\mk}_{0}(\SNnode{Edge})= \tuple{znd_1,znd_{23}}+\tuple{znd_4,znd_1}$, $\widehat{\mk}_{0}(\SNnode{Node} )= \tuple{znd_1+znd_{23}+znd_4}$
\end{center}
\noindent where all symbols (dynamic subclasses) refer to class $\class{N}$, and $|znd_1|=|znd_4| =1$, $|znd_{23}|= 2$. This symbolic marking represents six ordinary markings, including
$\mk_{G_0}$. A symbolic reachability graph (or SRG) is directly built from an initial symbolic marking, by means of a symbolic firing rule (and a canonical representative for SM). Skipping the technical details, a symbolic instance of  $\SNnode{R}_1$ folding the two bindings above is enabled in $\widehat{\mk}_{0}$; this symbolic instance may fire, leading to a new symbolic marking representing (among others) the ordinary markings $i)$ and $ii)$.

The SRG built (with the GreatSPN package) from $\widehat{\mk}_{0}$ is a quotient-graph of the RG, retaining liveness and safety properties: in the simple example we are considering, the SRG holds 9 nodes plus an absorbing one, each encoding a class of isomorphic graphs. When huge graphs are encoded with SN, the reduction achieved with the SRG in terms of generated states/arcs may be dramatic (e.g., a symbolic instance of transition
$\SNnode{R}_1$ may fold up to $|\class{N}|^3$ ordinary instances), even if bringing a SM to a canonical form is comparable to checking graph isomorphism.

\section{Exploiting SN Structural Analysis: an example}
\label{sec:struct-an}

In Section \ref{sec:rulecorr} we have established some conditions on
arc functions making a SN transition specify a well-defined graph rewriting rule. These conditions involve functional operators that can be solved in a fully automated/symbolic way through the  SNexpression tool, implementing the computation of a base set of structural properties \cite{Duth:93} directly on SN models,
without any unfolding. Each structural property may be expressed in terms of language \Lang\ , which is a small extension of arc functions.

Let us discuss now about the exploitation of these properties for validating rules, e.g., to figure out which rules of a GTS might concurrently apply.
Concurrent graph rewriting issues have been widely tackled in literature: we do not want to go into the details of a theoretical discussion, rather we aim at showing the potential of SN structural analysis in this field.

Symbolic structural relations are computed by properly combining arc functions through some operators: transpose, sum, difference, support, and composition.
A relation is a map $\mathcal{R}(t,t'): cd(t') \rightarrow 2^{cd(t)}$ that when applied to an instance $c'$ of $t'$ gives the set of instances of $t$ that are in such a relation with $(t',c')$.
Symbolic relations build on a couple of auxiliary ones, involving
a pair place/transition, both with arity $cd(p) \rightarrow 2^{cd(t)}$: 
 $\SfP{t}{p}= \overline{\Wminus{p}{t}}^{t}$ (\textit{Removed by}), given a color $c$ of $p$ provides the set of instances of $t$ that withdraw $c$ from $p$; $\AtP{t}{p}= \overline{\Wplus{p}{t}}^{t}$ (\textit{Added by}), given a color $c$ of $p$ provides the set of instances of $t$ that add $c$ to $p$.
Table \ref{tab:structrel} reports the definitions of base structural relations.

\emph{(Asymmetric) Structural Conflict}:
Two transition instances $(t,c)$ and $(t',c')$ are in conflict in a given marking $\mk$ if the firing of the former disables the latter. The structural conflict ($SC$) relation defines the necessary conditions that \textit{may} lead to an actual conflict in some marking. The symbolic relation $SC(t,t')$ maps a an instance $c'$ of $t'$ to the set of colour instances of $t$ that may disable $(t',c')$: this happens
either because $(t,c)$ withdraws a token from an input place which is shared by the two transitions, or because it adds a token into an output place which is connected to $t'$ through an inhibitor arc. These two cases are reflected in the $SC$ formula, which is is obtained by summing up over all shared input places and shared output-inhibitor places.
Observe that different instances of the {\em same} transition may be in conflict (auto conflict): the same expression can be used, but one must subtract from the set of conflicting instances the instance itself to which $SC$ applies (using the identity function).

\noindent
\emph{Structural Causal Connection}:
Two transition instances $(t,c)$ and $(t',c')$ are in causal connection if the firing of the former in a given marking $\mk$ causes the enabling of the latter. The structural causal connection ($SCC$) relation defines the necessary conditions that may lead to an actual causal connection in some marking. The symbolic relation $SCC(t,t')$, when applied to an instance $c'$ of $t'$, provides the set of instances $(t,c)$ that may cause the enabling of $(t',c')$. This happens if some output places of $t$ are input places for $t'$ and some input places of $t$ are inhibitor places for $t'$.

\noindent
\emph{Structural Mutual Exclusion}:
Two transition instances $(t,c)$ and $(t',c')$ are in (structural) mutual exclusion ($SME$) if
the enabling of $(t',c')$ in any $\mk$ implies that $(t,c)$ is not enabled, and viceversa.
This situation arises when a place $p$ does exist which is input for $t$ and inhibitor for $t'$, and the number of tokens (of any color) required in $p$ for the enabling of $t$ is greater than or equal to the upper bound on the number of tokens (of the same color) in $p$ imposed by the inhibitor arc connecting $p$ and $t'$.
The (symmetric) symbolic relation ${\it SME}(t,t')$
maps an instance $(t',c')$ to the set of instances of $t$ that are surely disabled in any marking where $(t',c')$ is enabled.
If all functions on input and inhibitor arcs were mappings onto sets
(i.e., on multisets with multiplicities $\leq 1$), as in the SN models
presented in this paper, 
then the ${\it SME}$ relation corresponds to the expression
in Table~\ref{tab:structrel}, that applies also when $t$ and $t'$ coincide\footnote{we refer to \cite{QEST2015} for a general treatment of SME}.

\paragraph{Application example} Structural relations can be used to validate the rules of a GTS formalized in terms of SN. In particular, it is possible to check which rules may concurrently apply, in the event a \textit{true concurrent} semantics were used. Using the structural calculus for SN we can -in a way, parametrically (i.e., symbolically) partition the set of instances of a given transition (rule) on the basis of a given relation with the instances of the other (or even the same) rule(s).

\begin{table}
\centering
\caption{Symbolic Structural relations in SN}
\label{tab:structrel}
\begin{tabular}{rcl}
\hline\noalign{\smallskip}
$SC(t,t')$   & = & $\bigcup_{p} \SfP{t}{p} \circ \overline{\In{t'}{p}} \ \  \cup\ $
 $ \AtP{t}{p} \circ \overline{\Inh{t'}{p}}$\\
 $SCC(t,t')$  & = & $\bigcup_{p} \AtP{t}{p} \circ \overline{\In{t'}{p}} \ \ \cup\ $
 $ \SfP{t}{p} \circ \overline{\Inh{t'}{p}} $\\
 ${\it SME}(t,t')$ & = & $\bigcup_{p } {\overline{\In{t}{p}}}^t  \circ \overline{\Inh{t'}{p}} \ \ \cup\ $ $ {\overline{\Inh{t}{p}}}^t \circ \overline{\In{t'}{p}}$ \\
\hline
\end{tabular}
\end{table}

In order to illustrate these concepts, let us consider the GTS in Figure \ref{fig:R1-3}. The two rules are potentially in conflict due to place \SNnode{Edge}, which is simultaneously an output place for one rule and an inhibitor place for the other. Instead, there are no potential conflicts due to the sharing of input places, since we can easily check that the expressions \SfP{t}{p} are null (by the way, a composition involving a null function results in $\emptyset$). As for the \textit{added by} expressions, we got the following non-null entries \footnote{all the calculus were done with SNExpression tool} (in the sequel, function supports are implicitly used):
\begin{align*}
\AtP{\SNnode{R}_1}{\SNnode{Edge}} =  \tuple{n_1,All,n_2} & & \AtP{\SNnode{R}_3}{\SNnode{Edge}}  =  \tuple{n_1}[n_1 = n_2]
\end{align*}
The first expression says that a color (token) $\tuple{c_1,c_2}$ may be pushed into place \SNnode{Edge} by \textit{any} instance of $\SNnode{R}_1$ (a triplet of colours) whose 1st and 3rd elements are equal to $c_1$ and $c_2$, respectively. The other expression says that a color $\tuple{c_1,c_2}$, with $c_1 = c_2$, may be pushed into place \SNnode{Edge} by the instance $\tuple{c_1}$ of  $\SNnode{R}_3$. Then, according with Table \ref{tab:structrel} we obtain:
\begin{align*}
SC(\SNnode{R}_1,\SNnode{R}_3)   =   \tuple{n_1,All,n_2} \circ  \tuple{n_1,All}  &  =  \tuple{n_1,All,All}  \\
SC(\SNnode{R}_3,\SNnode{R}_1)  =  \tuple{n_1}[n_1 = n_2] \circ \tuple{n_1,n_3} & = \tuple{n_1}[n_1 = n_3]  \\
\end{align*}
Again, the interpretation of these symbolic expressions is quite intuitive:
$SC(\SNnode{R}_1,\SNnode{R}_3)$ says that an instance $\tuple{c_1}$ of Rule
3 might be in conflict with (i.e., disabled by) any instance of Rule 1 having
color $c_1$ as first element; $SC(\SNnode{R}_3,\SNnode{R}_1)$ instead says that an instance $\tuple{c_1,c_2,c_3}$ of Rule
1, such that $c_1 = c_3$, might be in conflict with the instance $\tuple{c_1}$ of Rule 3.

The $SC$ relation, however, just outlines potential conflicts.
The previous outcome may be refined by computing $SME$: in fact, we observe that place $\SNnode{Edge}$ is both input and inhibitor for
$\SNnode{R}_1$, and inhibitor for $\SNnode{R}_3$. Then, according with Table \ref{tab:structrel} we obtain:
\begin{align*}
SME(\SNnode{R}_1,\SNnode{R}_3)   =   \tuple{All,n_1,All} + \tuple{n_1, All,All}  & &  SME(\SNnode{R}_3,\SNnode{R}_1) =  \tuple{n_1}+\tuple{n_2}  \\
\end{align*}
Notice that, according with the transpose rules and the relation's symmetry:  $SME(\SNnode{R}_3,\SNnode{R}_1)^t = SME(\SNnode{R}_1,\SNnode{R}_3) $. What is interesting, however, is that
$SC(\SNnode{R}_1,\SNnode{R}_3) \subset SME(\SNnode{R}_1,\SNnode{R}_3)$ and $SC(\SNnode{R}_3,\SNnode{R}_1) \subset SME(\SNnode{R}_3,\SNnode{R}_1)$, i.e., potentially conflicting instances of Rules 1 and 3 are in structural mutual exclusion. In other words, these two rules are potentially concurrent.

The same check may be done on instances of the \textit{same} rule. Consider $\SNnode{R}_1$: potential auto-conflicts due to place \SNnode{Edge} correspond to the symbolic expression:
\begin{align*}
SC(\SNnode{R}_1,\SNnode{R}_1) & =  \tuple{n_1,All-n_2,n_2} + \tuple{n_1,All-n_1,n_3}[n_1 = n_2] + \tuple{n_2,All,n_3}[n_1 \neq n_2]+\tuple{n_1,n_2,n_2}[n_2 \neq n_3]
\end{align*}
The mutually exclusive instances of the same transition correspond to the symbolic expression:
\begin{align*}
SME(\SNnode{R}_1,\SNnode{R}_1) & =  \tuple{n_2,All,n_3}+\tuple{n_1,All,n_2} + \tuple{n_1,n_3,All}+ \tuple{All,n_1,n_3}
\end{align*}
Also in this case, $SC(\SNnode{R}_1,\SNnode{R}_1) \subset SME(\SNnode{R}_1,\SNnode{R}_1)$, i.e., the instances of
$\SNnode{R}_1$ are potentially concurrent.
A similar check may be done for $\SNnode{R}_3$ instances.

In general, checking whether the rules of a GTS may concurrently take place (possibly identifying parametric concurrent subsets of rule instances) involves more complex calculations: think, e.g., of \textit{indirect conflicts} arising between non conflicting rule instances $(R,b)$ and $(R',b')$ enabled in marking \mk\ :  we fall in such a situation, e.g., if the firing of $(R,b)$ triggers a sequence of causally connected rule instances ending with an instance $(R'',b'')$ which is actually in conflict with (disables) $(R',b')$. Computing the \textit{transitive closure} of a structural relation \cite{QEST2015} is necessary to recognize indirect conflicts.  

\section{Conclusions and ongoing work}
\label{sec:concl}
We have presented a formalization of Graph Transformation Systems (GTS) based on Symmetric Nets (SN), a type of Coloured Petri nets featuring a particular syntax that outlines model symmetries. Each rule of a GTS corresponds to a transition of a SN which is properly connected to a couple of places encoding a graph. The advantages of this approach are numerous: we can exploit well established tools supporting the editing/analysis of SN, like the GreatSPN package; an operational interleaving semantics for GTS is provided in a natural way building the state-transition system of a SN; a compact state-transition system -called symbolic reachability graph, in which states (markings) representing isomorphic graphs are folded, can be directly derived once an initial symbolic graph encoding is set; some recent advances in SN (symbolic) structural analysis,
implemented in the SNExpression tool, may be exploited to check some conditions ensuring rule well-definiteness, to validate rules, and to check their potential concurrency;
in particular, a fully automated calculus of symbolic structural relations in SN models may be profitably used. All these concepts have been instantiated on a few, though significant, examples of graph rewriting rules, and a simple GTS. Throughout the paper we refer to the encoding of simple directed graphs.   

Ongoing work is in two main directions. The presented approach is general, we are therefore extending the class of encodable graphs to multigraphs (this extension is for free, it only requires that some well-definiteness conditions on rules are relaxed), bipartite graphs, hypergraphs, and so forth. Some SN features not used in the paper might be needed: for example (think of bi-or three-partite graphs), partitioning the colour class of nodes in two or more subclasses

A more theoretical research line involves a comparison of the SN based approach with classical approaches to GTS,
in particular the algebraic ones based on single/double pushout. We are firmly convinced that, under some quite general conditions, it is possible to characterize a SN rule as a pushout (in particular, a dpo) derivation. The practical implications of such a relationship (when confirmed) deserve further investigations. 

\nocite{*} 
\bibliographystyle{eptcs}
\bibliography{biblio}
\end{document}